# Superconductivity above 28 K in single unit cell FeSe films interfaced with GaO$_{2-\delta}$ layer on NdGaO$_3$(110)

Haohao Yang[1,*], Guanyu Zhou[1,*], Yuying Zhu[1,*], Guan-Ming Gong[1], Qinghua Zhang[2], Menghan Liao[1], Zheng Li[1], Cui Ding[1], Fanqi Meng[2], Mohsin Rafique[1], Heng Wang[1], Lin Gu[2,3], Ding Zhang[1,3,†], Lili Wang[1,3,†] and Qi-Kun Xue[1,3,†]

[1] State Key Laboratory of Low-Dimensional Quantum Physics, Department of Physics, Tsinghua University, Beijing 100084, China
[2] Laboratory for Advanced Materials & Electron Microscopy, Beijing National Laboratory for Condensed Matter Physics, Institute of Physics, Chinese Academy of Sciences, Beijing 100190, China
[3] Collaborative Innovation Center of Quantum Matter, Beijing 100084, China

[*]These authors contributed equally to this work.
[†]E-mail: dingzhang@tsinghua.edu.cn;liliwang@mail.tsinghua.edu.cn; qkxue@tsinghua.edu.cn

**Abstract**

We prepared single unit cell FeSe films on GaO$_{2-\delta}$ terminated perovskite NdGaO$_3$(110) substrates and performed *ex situ* transport and scanning transmission electron microscopy measurements on the FeTe protected films. Our experimental measurements showed that the single unit cell FeSe films interfaced with GaO$_{2-\delta}$ layer are electron doped via interface charge transfer. Most importantly, this type of heterostructure can host interface enhanced superconductivity with an onset temperature of about 28 K, which is much higher than the value of 8 K in bulk FeSe. Our work indicates that FeSe/GaO$_{2-\delta}$ can be a comparative platform with the FeSe/TiO$_{2-\delta}$ family for understanding the mechanism of interface enhanced high temperature superconductivity.

**Keywords:** FeSe/GaO$_{2-\delta}$, interface enhanced superconductivity, transport, scanning transmission electron microscopy

## 1. Introduction

The discovery of high temperature superconductivity in single unit cell (UC) FeSe on TiO$_{2-\delta}$ terminated perovskite SrTiO$_3$(001) substrates [1] has attracted intensive attention on searching for new superconducting systems with engineered interfaces as well as understanding the mechanism of interface high temperature superconductivity. In stark contrast to bulk FeSe—a superconductor with transition temperature $T_c \sim 8$ K at ambient pressure [2], the single UC FeSe on SrTiO$_3$(001) becomes superconducting at $T_c \sim 65$ K [3-5] or higher [6, 7], holding the record $T_c$ among all known Fe-based superconductors. The superconducting gap is enhanced by one order of magnitude compared to the value of bulk FeSe, i.e. 20 meV [1] vs. 2.2 meV [8]. What is more interesting is that monolayer FeSe on several TiO$_{2-\delta}$ terminated substrates, such as BaTiO$_3$[9], SrTiO$_3$(110) [10-12] and TiO$_2$ [13, 14], exhibit similarly high transition temperatures, despite the variation of lattice constant [15, 16]. The common features shared by all these 1-UC FeSe/TiO$_{2-\delta}$ systems are the enlarged electron pockets at the Brillouin zone (BZ) corners [3, 17, 18] and emergent replica bands 90-100 meV from the main bands [11, 14, 18]. They indicate that FeSe is electron-doped and the electrons therein couple with Ti-O stretching phonons [19, 20], respectively. On the other hand, electron-doped FeSe systems, such as K-coated multilayer FeSe, exhibit similar Fermi surfaces but universally lower $T_c$ (45 K) and smaller superconducting gaps (14 meV) [21-24]. This difference indicates that interface electron-phonon interaction plays an important role in further enhancing superconductivity by cooperatively mediating the cooper pairing [15, 16, 18, 22, 25].

The FeSe/TiO$_{2-\delta}$ interface bears sharp resemblance to the building blocks of cuprate and Fe-based high temperature superconductors. There, superconductivity emerges in the CuO$_2$ and FeAs layers, that are interfaced with oxide layers of BaO/SrO and LaO/SmO, respectively [26-28]. The fact that a variety of oxide layers can act as the charge reservoir layer motivates us to look for an oxide layer that is different from TiO$_{2-\delta}$. One recent endeavour by interfacing FeSe with MgO(001) proved that the onset superconducting temperature can be raised to 18 K [29]. Here, we report on the observation of superconductivity with an onset temperature of 28 K in single UC FeSe films epitaxially grown on perovskite NdGaO$_3$. Scanning transmission electron microscopy (STEM) reveals that FeSe sits on double GaO$_{2-\delta}$ in a striking similar manner as that of FeSe on TiO$_{2-\delta}$. Electron energy loss spectroscopy (EELS) further identifies the charge transfer across the FeSe/GaO$_{2-\delta}$ heterojunction.

## 2. Experimental

NdGaO$_3$ has been commonly used as a substrate for thin-film deposition of high temperature cuprate superconductors due to the matching lattice constants and thermal expansion coefficients [30, 31]. NdGaO$_3$ has a distorted orthorhombic structure with lattice parameters: a = 0.543 nm, b = 0.550 nm and c = 0.772 nm [32]. In order to achieve lattice matching between FeSe and NdGaO$_3$, we chose NdGaO$_3$(110). This crystal orientation corresponds to in-plane lattice constants: a$_{(110)}$ = 0.384 nm and b$_{(110)}$ = 0.386 nm (Fig. 1(a)), which are close to the in-plane lattice constant of FeSe: a$_{FeSe}$ = 0.379 nm [2]. To obtain single GaO$_{2-\delta}$-terminated surface, we pretreated NdGaO$_3$(110) substrates by wet chemical etching in a 10%-HCl solution for 45 min before thermal annealing in a tube furnace at 930°C under O$_2$ flux for 3 hours [33]. At higher annealing temperature or with longer annealing time, the GaO$_{2-\delta}$ layer partially decomposes such that both GaO$_{2-\delta}$ and NdO terminations occur. The epitaxial growth was carried out in an ultra-high vacuum (UHV) CreaTec molecular beam epitaxy (MBE) system. The base pressure was better than $1.0 \times 10^{-10}$ Torr. After being transferred into UHV chamber, NdGaO$_3$(110) substrates were heated at 600 °C under the Se flux for 30 min, and FeSe films were then grown by co-evaporating high-purity Fe (99.995 %) and Se (99.999 %) from standard Knudsen cells at a substrate temperature of 430 °C. The Fe/Se flux ratio was ~1:10 and the growth rate ~ 0.10 UC per minute. Then the samples were *in-situ* post-annealed at 450 °C to remove extra Se.

Since NdGaO$_3$ is highly insulating, we attached carbon nanotubes (tens of turns, ~ 2 mm wide) at one end of the substrate before loading the substrate to UHV chamber. This method guarantees *in-situ* room temperature scanning tunnelling microscopy (STM) morphology characterization of the overlaid FeSe film, indicating that carbon nanotubes electrically bridge the FeSe film and the contacting bar of the sample holder. Samples for *ex-situ* transport and STEM measurements were further capped by 10 UC FeTe films. The FeTe films were grown by co-evaporating Fe and Te from standard Knudsen cells at a substrate temperature of 255 °C. This capping method has been successfully used for protecting FeSe films against ambient contamination/oxidation [4, 10, 34].

STEM samples were prepared using Focused Ion Beam method. Cross-sectional lamella was thinned down to 100 nm thick at an accelerating voltage of 30 kV with a decreasing current from the maximum 2.5 nA, followed by fine polish at an accelerating voltage of 2 kV with a small current of 40 pA. STEM experiments were performed on an aberration-corrected ARM200CF (JEOL), operated at 200 keV and with double spherical aberration (Cs) correctors. The attainable resolution of the probe defined by the objective pre-field was 78 picometers. For the high angle annular dark field (HAADF) image, the collection angle was 90-370 mrad.

A closed-cycle system (Oxford Instruments TelatronPT) with base temperature of 1.6 K was used for transport measurements. Freshly cut indium cubes were cold pressed onto the samples as contacts. Samples were measured in a four-terminal configuration by employing the standard lock-in technique (1 μA at 13 Hz).

## 3. Results and discussion

FeSe forms crystalline films on Se-etched GaO$_{2-\delta}$ terminated NdGaO$_3$(110) substrates via a layer-by-layer mode. Figure 1(b) displays the morphology of FeSe films with a nominal coverage of ~ 1.0 UC on NdGaO$_3$ with both GaO$_{2-\delta}$ and NdO terminations and the corresponding line profile showing characteristic steps of 0.19 nm. Fig.1(c) shows 1.1 UC FeSe on single GaO$_{2-\delta}$ terminated substrate. This substrate hosts a step height of 0.38 nm. Clearly, FeSe films on GaO$_{2-\delta}$ are mostly homogeneous while films on NdO are fragmented islands with high density of sub-nm voids. FeSe films prepared on NdO-terminated surface show no superconducting transition. In the following, we focus on FeSe films on GaO$_{2-\delta}$ terminated NdGaO$_3$(110), as schematically shown in the bottom panel of Fig. 1(a).

STEM demonstrates that the interface consists of in-plane-expanded and out-of-plane compressed FeSe on double-GaO$_{2-\delta}$ layers. Figures 2(a) and 2(b) show typical atomically resolved high-angle annular dark field (HAADF) images of FeTe/1UC-FeSe/NdGaO$_3$(110) hetero-structures viewed along the [100] and [110] directions of FeSe, respectively, where Te, Se, Fe, Ga and Nd atoms appear as bright dots and are clearly identified with sharp contrast in intensity. On the NdGaO$_3$ side, Nd atoms are brighter than Ga due to larger atomic number. From bottom to top, a clear alternating sequence of Nd-Ga-Nd-Ga can be identified. Intriguingly, on top of these four layers and right beneath FeSe, there exists an extra layer of atoms (marked by empty circles). These atoms locate directly on top of Ga, but exhibit similar intensity and size. We speculate that the extra layer corresponds to GaO$_{2-\delta}$ layer, giving rise to a double-GaO$_{2-\delta}$ at the interface similar to that observed in FeSe/SrTiO$_3$ [34]. Between this extra layer and FeTe capping layer resides the interfacial Se-Fe-Se triple layer. The atomic resolution gives an averaged in-plane Se-Se distance of 0.384 $\pm$ 0.002 nm and an out-of-plane height of 0.134 $\pm$ 0.002 nm between the Fe layer and the bottom Se layer. The epitaxial films are therefore ~ 2% expanded in the plane and 7% compressed out-of-plane relative to bulk FeSe (0.379 nm and 0.145 nm) [2].

EELS measurements indicate that the interfacial single UC FeSe films get electron doped. Displayed in Figs. 2(c) is the EELS of Fe L$_{23}$-edge across a FeTe/6 UC-FeSe/NdGaO$_3$(110) hetero-structure. From the fifth to the second Fe layers, the intensity peaks of Fe L$_{23}$-edges don't show clear relative shift (the top four curves in Fig. 2(c)). In stark contrast, the intensity peaks of Fe L$_{23}$-edges in the first UC of FeSe that is directly interfaced with GaO$_{2-\delta}$ layer consistently get broadened and shift 0.4 $\pm$ 0.1 eV to higher energy, indicating that Fe ions therein accept electrons from the GaO$_{2-\delta}$ layer. Notably, such an Fe L$_{23}$-edge blue shift is observed at room temperature, whereas a blue shift of ~ 0.7 $\pm$ 0.1 eV was observed in 8 UC-FeSe/SrTiO$_3$ only at 10 K [35]. This contrast may stem from the distinctly different temperature dependent properties of SrTiO$_3$ and NdGaO$_3$ substrates. For example, the dielectric constant of SrTiO$_3$ increases by two orders of magnitude as the temperature decreases from room temperature to about 4 K, which does not happen in NdGaO$_3$ [36]. From the energy band perspective, NdGaO$_3$ has a larger band gap (3.8 eV) than SrTiO$_3$ and the band bending in NdGaO$_3$/SrTiO$_3$ heterostructure happens in a similar manner to that in LaAlO$_3$/SrTiO$_3$ [37]. These facts indicate that the electron affinity of NdGaO$_3$ is smaller than that of SrTiO$_3$. Therefore, the band bending in FeSe/NdGaO$_3$, which drives the charge transfer, is weaker than that in FeSe/SrTiO$_3$ [38].

Our most important finding is the observation of enhanced superconductivity with an onset temperature of 28 K in single UC FeSe on NdGaO$_3$(110). Displayed in Fig. 3 are the temperature dependent resistances of 1 UC, 3 UC and 4 UC FeSe samples. The schematic setup for transport measurements is shown in the inset. All these ultrathin FeSe films on NdGaO$_3$(110) substrates undergo superconducting transitions at 25-35 K. For the 1 UC sample, the resistance starts to drop at ~ 35 K; and for the 4 UC sample, the resistance starts to drop at ~ 30 K and reaches zero at 11 K. We define the onset superconducting temperature as the crossing point of the extrapolated lines from the normal state and the transition regime (gray lines). The upper inset panel in Fig. 3 summarizes the onset temperatures: $T_{onset}^{1\ UC} = 28.0$ K, $T_{onset}^{3\ UC} = 25.3$ K and $T_{onset}^{4\ UC} = 24.7$ K, all of which are more than three times the $T_c$ for bulk FeSe (8 K) [2]. Notably, the superconducting transition shifts to lower temperatures and widens considerably with increasing FeSe thicknesses-a typical feature of interface enhanced superconductivity. **For** conventional metal superconductors, however, a thicker film usually has a higher $T_c$ with a sharper transition. The opposite behaviour seen here reflects the decaying influence from the substrate. Previous transport measurements have revealed that thicker FeSe films on SrTiO$_3$ also exhibit lower transition temperatures as well as wider transition temperature regions, suggesting that only the first UC of FeSe at the interface is superconducting [35, 39]. The same thickness dependence seen in FeSe/NdGaO$_3$(110) system, therefore, indicates here also only the first UC FeSe becomes superconducting.

Figure 4 shows transport results of 1 UC and 4 UC FeSe samples under a perpendicular magnetic field. The transition broadens and shifts to lower temperatures at higher magnetic fields. For a 2D superconductor, its response to an external magnetic field normal to the 2D plane is expected to follow the linear Ginzburg-Landau formula [40]

$$\mu_0 H_{c2}(T) = \frac{\Phi_0}{2\pi\xi^2}\left(1 - \frac{T}{T_c}\right)$$

where $\mu_0 H_{c2}$ is the upper critical field, $\xi$ the in-plane coherent length at zero temperature, $\Phi_0 = h/2e$ the flux quantum and $T_c$ the superconducting transition temperature at zero field. Here, we define the transition temperature at each magnetic field as the temperature at which the resistance drops to 50% of the normal state resistance. As shown in the insets of Figs. 4(a) and 4(b), the linear fittings yield: $H_{c2,\ 1UC}(0) \sim 22.5$ T and $H_{c2,\ 4UC}(0) \sim 30.6$ T, respectively. Correspondingly, the in-plane coherence lengths are $\xi_{1UC} = 3.59$ nm and $\xi_{4UC} = 3.28$ nm. Notably, data points of 4 UC FeSe exhibit a rapid increase in temperature when approaching zero field, presumably due to phase separation.

The magneto transport data can be further used to estimate the London penetration depth, a key parameter for a superconductor. To this end, we convert data in Figs. 4(a) and 4(b) to Arrhenius plots shown in Figs. 4(c) and 4(d). The straight lines here indicate activation of vortices in the presence of a magnetic field. We then apply the formula $R(T, H) = R_0(H) exp[-U_0(H)/T]$, to extract the activation energy $U_0(H)$ [41]. As displayed in the corresponding insets, $U_0(H)$ scales roughly linearly with $\ln\mu_0 H$, likely resulting from collective creeping of vortices [41]. The slope of this linear trend can therefore be employed to estimate the penetration depth as: $-\frac{dU_0(H)}{d\ln\mu_0 H} = \frac{\Phi_0^2 d}{256\pi^3 \Lambda^2}$, where $\Lambda$ the London penetration depth. By assuming $d = 0.55$ nm (Only interfacial single UC FeSe is superconducting), we obtain London penetration depths: $\Lambda_{1\ UC}= 84$ nm and $\Lambda_{4\ UC}= 126$ nm. The Ginzburg-Landau parameter $\kappa=\Lambda/\xi$ is therefore larger than 20, indicating a type II superconductor.

NdGaO$_3$ is the third type of oxides, after TiO$_{2-\delta}$-family (including SrTiO$_3$, BaTiO$_3$, TiO$_2$ etc.) and MgO layers, that can enhance superconductivity in single UC FeSe films. The $T_c$ ranks between the two previous ones [4, 29]. Similarly, interface charge transfer is identified from EELS. In comparison to FeSe on SrTiO$_3$(001), FeSe on NdGaO$_3$ possesses smaller in-plane lattice constants and larger out-of-plane Fe-Se distance. The band bending and screening effect are also weaker (estimated from work functions and dielectric constants, respectively). They correlate with the observed lower $T_c$, suggesting that superconductivity in FeSe indeed depends on the in-plane stretching [42] and band bending [21-24]. Our work also hints at a universal method to enhanced superconductivity by fabricating FeSe/oxide interface. The key technical point is to achieve atomically flat and uniform oxide layer to guarantee atomically sharp interfaces. A practical approach is to prepare high quality oxide layers by using oxide-MBE or pulse laser deposition techniques. From the perspective of growth kinetics, multi-heterostructures of FeSe/GaO$_{2-\delta}$ with sharp interfaces could be more easily achievable than FeSe/TiO$_{2-\delta}$, because gallium oxides could be grown at lower temperature, for Ga has higher saturated vapour pressure than Ti. Fabricating multi-layers of FeSe/GaO$_{2-\delta}$ may be a feasible way to further enhance superconductivity.

## 4. Summary

We observe interface enhanced superconductivity in single UC FeSe films grown on GaO$_{2-\delta}$-terminated NdGaO$_3$(110) substrates. Similar to FeSe/TiO$_{2-\delta}$ interface, we identify interface charge transfer. Our finding suggests that FeSe/GaO$_{2-\delta}$ is a new platform for investigating the mechanism of interface high temperature superconductivity. This work could also facilitate fabrication of sandwiched oxide/FeSe/oxide hetero-structures to achieve higher $T_c$.

**Acknowledgements**

This work was supported by the National Natural Science Foundation of China (11574174, 11774193, 11790311, 11404183, 51522212, 51421002, and 51672307), the National Basic Research Program of China (2015CB921000 and 2014CB921002), and the Strategic Priority Research Program of Chinese Academy of Sciences (Grant No. XDB07030200).

**Conflict of interest**

The authors declare that they have no conflict of interest.

**References**

[1] Wang Q-Y, Li Z, Zhang W-H, et al. Interface-induced high-temperature superconductivity in single unit-cell FeSe flms on SrTiO$_3$. Chin Phys Lett 2012; 29:037402.
[2] Hsu FC, Luo JY, Yeh KW, et al. Superconductivity in the PbO-type structure α-FeSe. Proc Natl Acad Sci USA 2008; 105:14262-14264.
[3] He S, He J, Zhang W, et al. Phase diagram and electronic indication of high-temperature superconductivity at 65 k in single-layer FeSe films. Nature Mater 2013; 12:605-610.
[4] Zhang W-H, Sun Y, Zhang J-S, et al. Direct observation of high-temperature superconductivity in one-unit-cell FeSe films. Chin Phys Lett 2014; 31:017401.
[5] Zhang Z, Wang Y-H, Song Q, et al. Onset of the Meissner effect at 65 K in FeSe thin film grown on Nb-doped SrTiO$_3$ substrate. Sci Bull 2015; 60:1301-1304.
[6] Deng LZ, Lv B, Wu Z, et al. Meissner and mesoscopic superconducting states in 1–4 unit-cell FeSe films. Phys Rev B 2014; 90:214513.
[7] Ge JF, Liu ZL, Liu C, et al. Superconductivity above 100 k in single-layer FeSe films on doped SrTiO$_3$. Nature Mater 2015; 14:285-289.
[8] Song C-L, Wang Y-L, Cheng P, et al. Direct observation of nodes and twofold symmetry in FeSe superconductor. Science 2011; 332:1410-1413.
[9] Peng R, Xu HC, Tan SY, et al. Tuning the band structure and superconductivity in single-layer FeSe by interface engineering. Nat Commun 2014; 5:5044.
[10] Zhou G, Zhang D, Liu C, et al. Interface induced high temperature superconductivity in single unit-cell FeSe on SrTiO$_3$(110). Appl Phys Lett 2016; 108:202603.
[11] Zhang C, Liu Z, Chen Z, et al. Ubiquitous strong electron–phonon coupling at the interface of FeSe/SrTiO$_3$. Nature Commun 2017; 8:14468.
[12] Zhang P, Peng XL, Qian T, et al. Observation of high-tc superconductivity in rectangular FeSe/SrTiO$_3$(110) monolayers. Phys Rev B 2016; 94:104510.
[13] Ding H, Lv YF, Zhao K, et al. High-temperature superconductivity in single-unit-cell FeSe films on anatase TiO$_2$(001). Phys Rev Lett 2016; 117:067001.
[14] Rebec SN, Jia T, Zhang C, et al. Coexistence of replica bands and superconductivity in FeSe monolayer films. Phys Rev Lett 2017; 118:067002.
[15] Wang L, Ma X-C, Xue Q-K. Interface high-temperature superconductivity. Supercond Sci Technol 2016; 29: 123001.
[16] Wang L, Xue Q-K. High temperature superconductivity in single-unit-cell-FeSe/TiO$_{2-\delta}$. AAPPS Bulletin 2017; 27:
[17] Tan S, Zhang Y, Xia M, et al. Interface-induced superconductivity and strain-dependent spin density waves in FeSe/SrTiO$_3$ thin films. Nature Mater 2013; 12:634-640.
[18] Lee JJ, Schmitt FT, Moore RG, et al. Interfacial mode coupling as the origin of the enhancement of tc in FeSe films on SrTiO$_3$. Nature 2014; 515:245-248.
[19] Xie Y, Cao HY, Zhou Y, et al. Oxygen vacancy induced flat phonon mode at FeSe/SrTiO$_3$ interface. Sci Rep 2015; 5:10011.
[20] Zhang S, Guan J, Jia X, et al. Role of SrTiO$_3$ phonon penetrating into thin FeSe films in the enhancement of superconductivity. Phys Rev B 2016; 94:081116.
[21] Miyata Y, Nakayama K, Sugawara K, et al. High-temperature superconductivity in potassium-coated multilayer FeSe thin films. Nature materials 2015; 14:775-779.
[22] Tang C, Liu C, Zhou G, et al. Interface-enhanced electron-phonon coupling and high-temperature superconductivity in potassium-coated ultrathin FeSe films on SrTiO$_3$. Phys Rev B 2016;93:020507.
[23] Zhang WH, Liu X, Wen CH, et al. Effects of surface electron doping and substrate on the superconductivity of epitaxial FeSe films. Nano Lett 2016; 16:1969.
[24] Wen CH, Xu HC, Chen C, et al. Anomalous correlation effects and unique phase diagram of electron-doped FeSe revealed by photoemission spectroscopy. Nature Commun 2016; 7:10840.
[25] Lee D-H. Routes to high-temperature superconductivity: A lesson from FeSe/SrTiO$_3$. Annual Review of Condensed Matter Physics 2018; 9:261-282.
[26] Chu CW, Deng LZ, Lv B. Hole-doped cuprate high temperature superconductors. Physica C: Superconductivity and its Applications 2015; 514:290-313.
[27] Kamihara Y, Watanabe T, Hirano M, et al. Iron-based layered superconductor La[O$_{1-x}$F$_x$]FeAs (x = 0.05-0.12) with $T_c$ = 26 K. J Am Chem Soc 2008; 130:3296.
[28] Ren Z, Lu W, Yang J, et al. Superconductivity at 55 K in iron-based f-doped layered quaternary compound Sm[O$_{1-x}$F$_x$]FeAs. Chin Phys Lett 2008; 25:2215.

[29] Zhou G, Zhang Q, Zheng F, et al. Interface enhanced superconductivity in monolayer FeSe films on MgO(001): Charge transfer with atomic substitution. Sci Bull 2018; 63:747.

[30] Koren G, Gupta A, Giess EA, et al. Epitaxial films of $YBa_2Cu_3O_{7-\delta}$ on $NdGaO_3$, $LaGaO_3$, and $SrTiO_3$ substrates deposited by laser ablation. Appl Phys Lett 1989; 54:1054.

[31] Phillips JM. Substrate selection for high-temperature superconducting thin films. J Appl Phys 1996; 79:1829.

[32] Krivchikov AI, Gorodilov BY, Kolobov IG, et al. Structure, sound velocity, and thermal conductivity of the perovskite $NdGaO_3$. Low Temperature Physics 2000; 26:370.

[33] Leca V, Blank DHA, Rijnders G. Termination control of $NdGaO_3$ crystal surfaces by selective chemical etching. arXiv:12022256 2012.

[34] Li F, Zhang Q, Tang C, et al. Atomically resolved $FeSe/SrTiO_3$(001) interface structure by scanning transmission electron microscopy. 2D Mater 2016; 3:024002.

[35] Zhao W, Li M, Chang C-Z, et al. Direct imaging of electron transfer and its influence on superconducting pairing at $FeSe/SrTiO_3$ interface. Sci Adv 2018; 4:2682.

[36] Krupka J, Geyer R G, Kuhn M, et al. Dielectric properties of single crystals of $Al_2O_3$, $LaAlO_3$, $NdGaO_3$, and MgO at cryogenic temperatures. IEEE Trans Microw Theory Tech 1994; 42:1886.

[37] Treske U, Heming N, Knupfer M, et al. Universal electronic structure of polar oxide hetero-interfaces. Sci Rep 2015; 5:14506.

[38] Zhang H, Zhang D, Lu X, et al. Origin of charge transfer and enhanced electron–phonon coupling in single unit-cell FeSe films on $SrTiO_3$. Nat Commun 2017; 8:214.

[39] Wang Q, Zhang W, Zhang Z, et al. Thickness dependence of superconductivity and superconductor–insulator transition in ultrathin FeSe films on $SrTiO_3$(001) substrate. 2D Materials 2015; 2:044012.

[40] Kozuka Y, Kim M, Bell C, et al. Two-dimensional normal-state quantum oscillations in a superconducting heterostructure. Nature 2009; 462:487.

[41] Ephron D, Yazdani A, Kapitulnik A, et al. Observation of quantum dissipation in the vortex state of a highly disordered superconducting thin film. Phys Rev Lett 1996; 76:1529.

[42] Fukaya Y, Zhou G, Zheng F, et al. Asymmetrically optimized structure in a high-$T_c$ single unit-cell FeSe superconductor. J Phys: Condens Matter 2019; 31:055701.

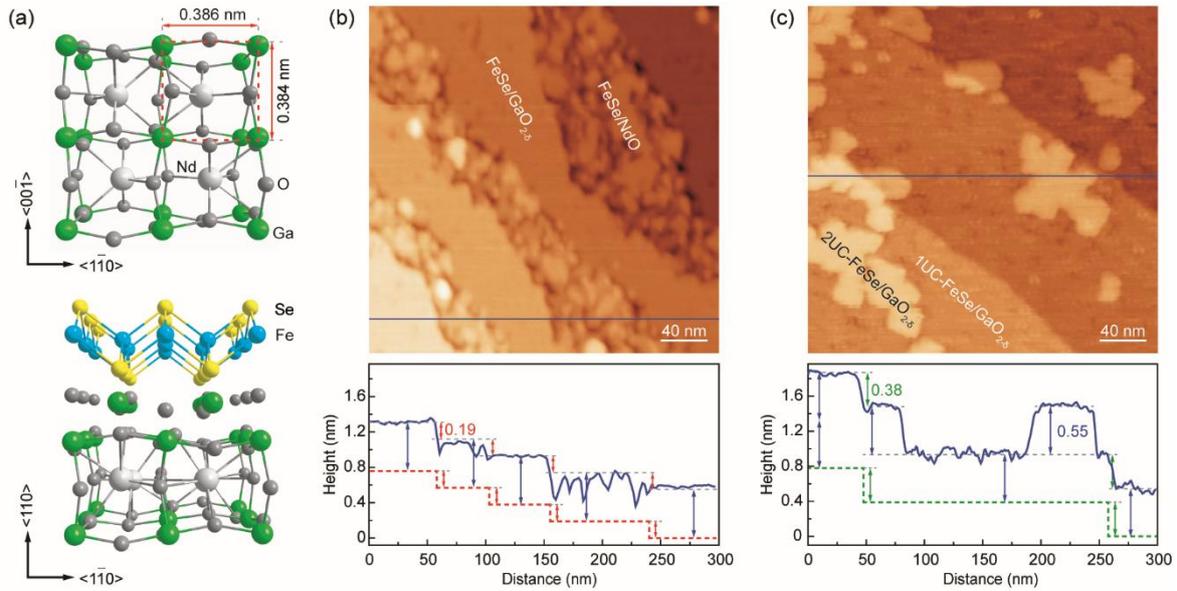

**Figure 1.** Structure and morphologies of FeSe films on $GaO_{2-\delta}$-terminated $NdGaO_3(110)$ substrates. (a) Schematic structure of $GaO_{2-\delta}$ terminated $NdGaO_3(110)$ surface (top panel) and 1 UC FeSe/$NdGaO_3$ hetero-structure (bottom panel). (b) and (c) Typical STM images of FeSe films on $GaO_{2-\delta}$ and NdO co-existed (sample bias $V_s$ =2.8 V, tunneling current $I_t$ =310 pA) and single $GaO_{2-\delta}$ terminated $NdGaO_3(110)$ ($V_s$ =2.0 V, $I_t$ =350 pA) substrates, respectively. Line profiles (lower panels) correspond to the blue lines in the morphology images. The dashed lines represent the stepped surfaces of $NdGaO_3(110)$ substrates with red/green double arrows mark steps of 0.19/0.38 nm high. The blue double arrows indicate the height of 1 UC FeSe.

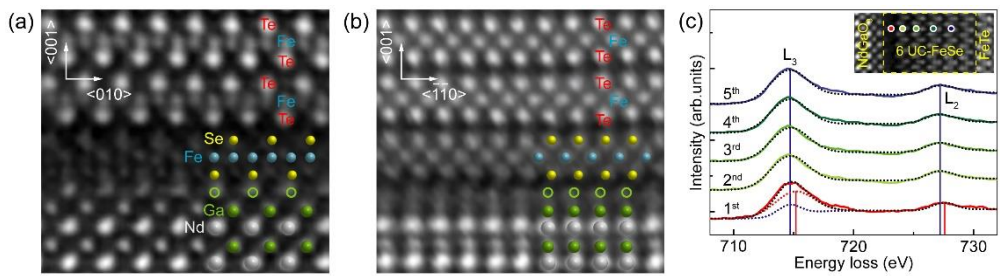

**Figure 2.** STEM characterization of FeTe/FeSe/$NdGaO_3(110)$ hetero-structures. (a) and (b) HAADF images of FeTe/1 UC-FeSe/$NdGaO_3(110)$ hetero-structure viewed along [100] and [110] directions of FeSe, respectively. (c) the EELS of Fe $L_{23}$-edge across a FeTe/6 UC-FeSe/$NdGaO_3(110)$ hetero-structure. The dots shown in the insert HAADF images mark the points where the spectra were taken. Dotted curves are Gaussian fitted spectra. For the $L_3$ edge in the first UC FeSe, the black dotted curve corresponds to the sum of the blue and the red Gaussian fitting curves. The Vertical lines are guides to the eye.

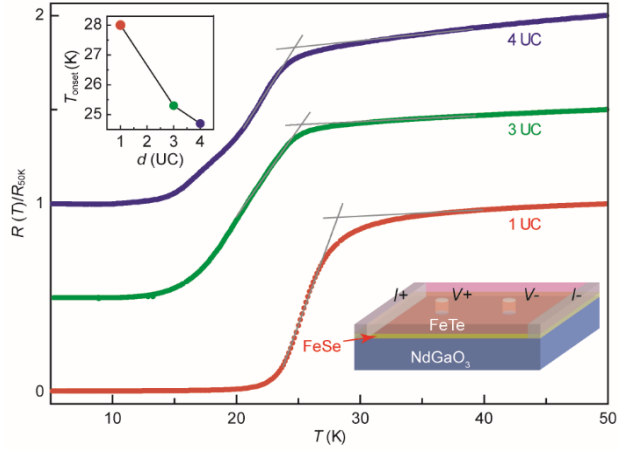

**Figure 3.** The temperature dependence of resistance (normalized at 50 K) of FeTe/FeSe/NdGaO$_3$(110) hetero-structures under zero magnetic field. Curves are vertically offset for clarity. Lower inset panel shows the schematic for the four-probe transport measurement, and upper inset panel the superconducting transition temperature-thickness relation.

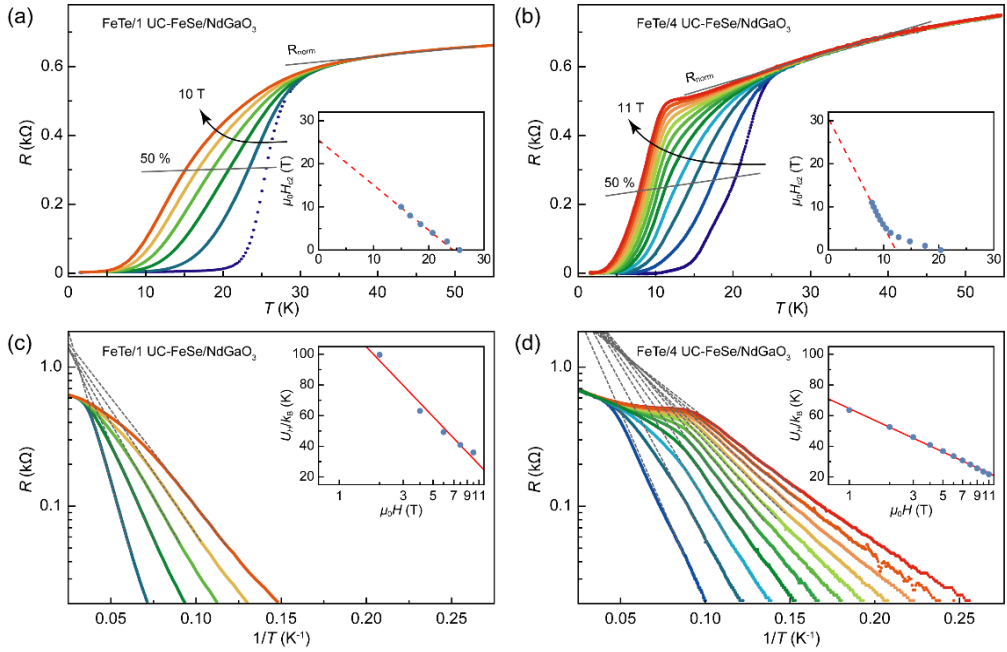

**Figure 4.** Magneto-transport of FeTe/FeSe/NdGaO$_3$(110) hetero-structures. (a) and (b) The temperature dependent resistance under various perpendicular magnetic fields for FeTe/FeSe/NdGaO$_3$(110) hetero-structures that consist of 1 UC and 4 UC FeSe films, respectively. The insets show the upper critical field as a function of temperature and the corresponding linear fittings. (c) and (d) the corresponding Arrhenius plots of (a) and (b), respectively. Gray dashed lines show linear fits in the low-temperature regime, reflecting the thermal activation behavior. Inset shows the activation energy $U_H/k_B$ as a function of the perpendicular magnetic field with a linear fitting marked by red solid line.